%% file: Manuscript.tex
\RequirePackage{fixltx2e}
\documentclass[pre,reprint,amssymb,amsmath,amsfonts,aps,floatfix,superscriptaddress]{revtex4-1}
\usepackage{graphicx, multirow, rotating, units, upgreek, xspace}

\newcommand{\Epi}{\affiliation{Department of Epileptology, University of Bonn, Sigmund-Freud-Stra{\ss}e~25, 53105~Bonn, Germany}}
\newcommand{\HISKP}{\affiliation{Helmholtz Institute for Radiation and Nuclear Physics, University of Bonn, \\Nussallee~14--16, 53115~Bonn, Germany}}
\newcommand{\IZKS}{\affiliation {Interdisciplinary Center for Complex Systems, University of Bonn, Br{\"u}hler Stra{\ss}e~7, 53175~Bonn, Germany}}

\newcommand{\kl}[1]{\left( #1 \right)}
\newcommand{\klg}[1]{\left\{ #1 \right\}}

\newcommand{\defi}{\mathrel{\mathop:}=}

\begin{document}

\title{Evolving networks in the human epileptic brain}

\author{Klaus Lehnertz}
\email{klaus.lehnertz@ukb.uni-bonn.de}
\Epi \HISKP \IZKS

\author{Gerrit Ansmann}
\Epi \HISKP \IZKS

\author{Stephan Bialonski}
\Epi \HISKP \IZKS

\author{Henning Dickten}
\Epi \HISKP \IZKS

\author{Christian Geier}
\Epi \HISKP

\author{Stephan Porz}
\Epi \HISKP

\begin{abstract}
Network theory provides novel concepts that promise an improved characterization of interacting dynamical systems.
Within this framework, evolving networks can be considered as being composed of nodes, representing systems, and of time-varying edges, representing interactions between these systems.
This approach is highly attractive to further our understanding of the physiological and pathophysiological dynamics in human brain networks.
Indeed, there is growing evidence that the epileptic process can be regarded as a large-scale network phenomenon.
We here review methodologies for inferring networks from empirical time series and for a characterization of these evolving networks.
We summarize recent findings derived from studies that investigate human epileptic brain networks evolving on timescales ranging from few seconds to weeks. We point to possible pitfalls and open issues, and discuss future perspectives.
\end{abstract}

\maketitle

\section{Introduction}
Over the past decade, network theory has contributed significantly to improving our understanding of spatially extended, complex dynamical systems, with wide applications in diverse fields, ranging from physics to biology and medicine \cite{Strogatz2001, Albert2002, Dorogovtsev2002, Newman2003, Boccaletti2006a, Reijneveld2007, Arenas2008, Bullmore2009, Fortunato2010, Barabasi2011, Barthelemy2011, Sporns2011a, Bashan2012, Newman2012, Stam2012b}.
The human brain is an open, dissipative, and adaptive dynamical system, which can be regarded as a network of interacting subsystems.
Due to its complex structure, its immense functionality, and --~as in the case of brain pathologies~-- due to the coexistence of normal and abnormal functions and/or structures, the brain can be regarded as one of the most complex and fascinating systems in nature.
The neocortex of human --~a thin, extended, convoluted sheet of tissue with a surface area of approx.~$\unit[2600]{cm^2}$, and thickness \unit[3--4]{mm} \cite{Braitenberg1991, Murre1995}~-- contains up to $10^{10}$ neurons, which are connected with each other and with cells in other parts of the brain by about $10^{12}$ synapses~\cite{Mountcastle1997}.
The length of all connections amounts to \unit[$10^7$--$10^9$]{m}.
The highly interconnected networks in the brain can generate a wide variety of synchronized activities, including those underlying epileptic seizures, which often appear as a transformation of otherwise normal brain rhythms.

With 50~million affected individuals worldwide \cite{Duncan2006, Guerrini2006}, epilepsy represents one of the most common neurological disorders~\cite{Engel1997}, second only to stroke.
Epilepsy is defined as \textit{a~disorder of the brain characterized by an enduring predisposition to generate epileptic seizures and by the neurobiologic, cognitive, psychological, and social consequences of this condition}~\cite{Fisher2005}.
For about \unit[30]{\%} of epilepsy patients, seizures remain poorly controlled despite maximal medical management \cite{Schuele2008, Spencer2008, Percurra2011, deTisi2011}.
There is thus a strong need for new curative treatments \cite{Morrell2006, Stacey2008}.

An epileptic seizure is defined as \textit{a transient occurrence of signs and/or symptoms due to abnormal excessive or synchronous neuronal activity in the brain} \cite{Fisher2005, Engel2006}.
Epileptic seizures may be accompanied by an impairment or loss of consciousness, psychic, autonomic or sensory symptoms, or motor phenomena.
Generalized-onset seizures are believed to instantaneously involve almost the entire brain~\cite{Engel2006}, while focal-onset seizures appear to originate from a circumscribed region of the brain \textit{(epileptic focus} \cite{Rosenow2001, Kahane2006}).
These simplistic concepts of focal and generalized seizures, however, are being challenged by increasing evidence of seizure onset within a network of brain regions \textit{(epileptic network)} \cite{Bertram1998, Bragin2000, Spencer2002, Lemieux2011}. This supports a new approach to classification of seizures and epilepsies~\cite{Berg2011}.

The concept of an epileptic network comprises anatomically, and more importantly, functionally connected cortical and subcortical brain structures and regions.
Since the timescale between onset and offset of a seizure is orders of magnitude smaller than that of any plausible change in the underlying structural components (such as neurons, axons or dendrites), seizures (and other related pathophysiological dynamics) may emerge from, may spread via, and may be terminated by network constituents that generate and sustain normal, physiological brain dynamics during the seizure-free interval.

Understanding the emergence of epilepsy and seizures from epileptic brain networks calls for approaches that take into account the interplay between the dynamic properties of nodes and the network structures connecting them.
When investigating epileptic brain networks, nodes are usually assumed to represent distinct brain regions and edges represent interactions between them, and these nodes and edges constitute a \textit{functional network.}
Epileptic brain networks are \textit{evolving functional networks} since their edges may change on various timescales, depending on physiological and pathophysiological conditions.

In this review, we summarize recent conceptual and methodological developments that aim at an improved inference and characterization of evolving epileptic brain networks.
We highlight areas that are under active investigation and that promise to provide new insights into the complex spatial and temporal dynamics of these networks.
We review frequently used approaches to infer functional networks from multichannel recordings of neural activities (Section~\ref{sec:inferring_networks}) as well as network and node characteristics that are most commonly used for  investigating epileptic brain networks (Section~\ref{sec:characterizing_networks}).
In Section~\ref{sec:epinetworks} we summarize findings obtained from studies that aim at characterizing evolving epileptic brain networks with respect to various physiological and pathophysiological conditions.
Finally, in Section~\ref{sec:conclusion} we draw our conclusions and give an outlook.

\section{Inferring functional brain networks}
\label{sec:inferring_networks}
Functional brain networks are supposed to reflect the interaction dynamics between brain regions.
Representing the complex system brain as a network, however, requires identification of nodes and edges.
This is a challenging issue given the complex structural and functional organization of the brain --~from the level of single neurons via microcolumns (containing some tens of neurons) and macrocolumns (consisting of some tens of microcolumns) to the level of brain regions, lobes, and functional brain systems~-- as well as methodological limitations in assessing this organization \cite{Ioannides2007, Antiqueira2010, Bialonski2010, Hayasaka2010, Zalesky2010, Bialonski2011b, Gerhard2011, Kaiser2011, Bialonski2012}.
Brain regions (nodes) are usually associated with sensors that are placed to sufficiently capture the node dynamics.
When characterizing edges, one is faced with the problem that the underlying equations of motion are not known and that interactions between brain regions cannot be measured directly.
Thus, usually time series analysis techniques are employed to quantify linear or nonlinear interdependencies between observables of brain regions.

\subsection{Acquiring time series of neural activity}
\label{sec:acquiring_time_series}
There are currently three recording techniques that are mainly used to obtain time series of neural activity, namely electroencephalography (EEG), magnetoencephalography (MEG), and functional magnetic resonance imaging (fMRI).
Each of these techniques assesses different aspects of neuronal activity and has its own spatial and temporal resolution as well as its way of associating brain regions to network nodes.

With EEG~\cite{Nunez2006} and MEG~\cite{Hamalainen1993}, electric and magnetic correlates of neural activities outside the head are measured with sensors that are placed according to standard schemes.
In some epilepsy patients undergoing presurgical evaluation~\cite{Rosenow2001}, sensors are placed intracranially, which allows for directly recording neural activities from within deeper brain structures and from the surface of the brain (iEEG)~\cite{Engel2005}.
In the following we use EEG for both, surface and intracranial EEG.
For all recording techniques, volume conduction and dense spatial sampling can give rise to mostly unavoidable influences like transitivity and common sources (see Section \ref{sec:estimating_interactions}), which need to be addressed in subsequent analysis steps.
For EEG the recording montage together with the choice of a reference electrode is a notoriously ill-defined problem \cite{Hagemann2001, Yao2005, Hu2010}.
An important advantage of EEG is the ability to perform recordings over extended periods of time (days to weeks), such that a wide spectrum of physiological and pathophysiological activities can be captured.
EEG and MEG sample brain activities with a time resolution of a few milliseconds, and sensor placement limits spatial resolution besides the mentioned influences.

With fMRI~\cite{Logothetis2008} neural activity is assessed indirectly via associated changes in blood oxygenation.
While this can be captured with very high spatial resolution, the temporal resolution is orders of magnitude lower than with EEG or MEG.

\subsection{Estimating interactions from time series}
\label{sec:estimating_interactions}
A plethora of analysis techniques is available to estimate strength and direction of interactions from time series.
These estimators originate from synchronization theory, nonlinear dynamics, information theory, statistical physics, and from the theory of stochastic processes (for an overview, see Refs.~\cite{Pikovsky_Book2001, Kantz2003, Pereda2005, Hlavackova2007, Marwan2007, Lehnertz2009b, Friedrich2011, Lehnertz2011}).
Here we highlight some of the more recent developments and improvements.

When analyzing interactions between several systems, one may be faced with the problem of \textit{transitivity:}
many estimators do not allow for distinguishing between direct and indirect interactions \cite{Bialonski2010, Zalesky2012} and therefore spurious edges between network nodes may be inferred.
This issue has been addressed through the use of partialization techniques \cite{Langford2001, Vakorin2009, Jamsek2010, Nawrath2010, Jalili2011, Hlinka2012, Sommerlade2012} but their suitability for analyses of empirical data remains to be shown.
Another frequently arising difficulty is due to the problem of \textit{common sources} \cite{Bialonski2010, Gerhard2011}:
sensors which are spatially close are likely to pick up very similar activities.
This can lead to spuriously high estimates of strengths of interactions but can probably be avoided using more advanced estimators for phase synchronization \cite{Vinck2011, Stam2012}.
Other developments that promise to further advance characterization of interactions include an optimized mixed state-space embedding \cite{Vlachos2010}, improved phase determination \cite{Kim2011, Wacker2011, Wacker2011b, Schwabedal2012, Stankovski2012}, bivariate surrogates \cite{Romano2009, Andrzejak2011, Papana2011, Rummel2011}, usage of ranks for nonlinear interdependencies~\cite{Chicharro2009}, cross-frequency decomposition~\cite{Nikulin2012}, improved recurrence estimators~\cite{Schultz2011}, multivariate and delayed information transfer \cite{Overbey2009, Ito2011, Pompe2011, Shibuya2011, Runge2012}, and approaches that characterize interactions even for transient dynamics \cite{Wagner2010a, Hempel2011, Martini2011}.
We note that up to now there are no commonly accepted approaches to estimate interactions and their properties from time series.

The human brain is certainly a non-stationary system, but for most estimators at least approximate stationarity is required.
Therefore it is advisable to perform a time-resolved analysis, which is carried out via a sliding-window approach.
A trade-off has to be made between approximate stationarity and the required statistical accuracy for the calculation of the estimator.
Typically, windows spanning several tens of seconds of brain activity are assumed to be acceptable \cite{Blanco1995, Rieke2003, Dikanev2005}.
For each of these windows, an interaction matrix $I$ containing estimates of interactions between all pairs of $n$ sampled brain regions is obtained.
The entry $I_{ij}$ denotes the estimate of an interaction property between nodes $i$ and $j$ ($i,j \in \klg{1, \ldots, n}$) of a network.

\subsection{Constructing functional networks}
\label{sec:constructing_networks}
From the interaction matrix, binary or weighted as well as undirected or directed networks can be derived.
However, only few methods for directed functional networks exist and most network characteristics rely on (weighted) undirected networks.

An \textbf{undirected binary network} can be represented by a symmetric \textit{adjacency matrix} $A \in \klg{0,1}^{n \times n}$ whose entry $A_{ij}$ is 1 if there is an edge between nodes $i$ and $j$, and 0~otherwise.
A commonly used approach is to consider these nodes as connected if the strength of interaction $I_{ij}$ exceeds some \textit{threshold}~$T$:
\[ A_{ij} = \begin{cases} 1 &\text{if } I_{ij} > T \\ 0 & \text{else} \end{cases}. \]
The value of $T$ is either chosen fixed \cite{Kramer2008, Meunier2009, VandenHeuvel2009} or properties of the resulting functional networks are investigated over a range of values of $T$ \cite{Horstmann2010, vanWijk2010}.
The threshold $T$ may also be determined adaptively \cite{Bassett2006, Zanin2012}, e.g., by choosing its maximal value such that the resulting network is not unconnected~\cite{Schindler2008a}.
Another strategy is to consider an edge to exist if the corresponding interaction is significant according to some statistical test \cite{Donges2009, Kramer2009, Rubinov2009, Emmert-Streib2010b}.

An \textbf{undirected weighted network} can be described by a weight matrix $W \in \mathbb{R}_+^{n \times n}$.
Often, all edges are considered to exist, such that $W$ fully describes a network and no adjacency matrix $A$ needs to be taken into account.
For weighted networks, we will restrict ourselves to such \textit{complete} networks in the following.
The easiest way to derive the weight matrix from the interaction matrix is to let the edge weights be identical to the respective strengths of interaction: $W_{ij} = I_{ij}$.
In order to eliminate the influence of the mean strength of interaction ($\bar I = 2 \sum_{i=1}^n \sum_{j=1}^{i-1} I_{ij} / \kl{n\kl{n-1}}$) on the resulting functional network, the mean weight may be set to~1:
\[ W_{ij} = I_{ij} - \bar I + 1 \text{\quad or\quad} W_{ij} = I_{ij} / \bar I. \]
Most normalizations of network characteristics also eliminate this influence, e.g., most null models (see Section~\ref{nullmodel}) constrain the mean edge weight.
The influence of all properties of the distribution of the estimated strengths of interactions can be eliminated by assigning weights from a given distribution using ranks~\cite{Kuhnert2012}.
Again, edge weights can be considered to be~0 (or edges to be non-existent), if the respective interaction estimate is not significant~\cite{Chavez2010}.

\section{Characterizing networks}
\label{sec:characterizing_networks}
To characterize functional networks, methods from graph theory are employed.
Some of these methods are based on concepts that have been developed and used since the 1970s for social network analysis and have since been refined and proven worthy as an important tool for understanding networks in various scientific fields; others have been developed only recently.
While we restrict ourselves to those characteristics that are most commonly used for the analysis of epileptic brain networks, a plethora of other characteristics and modifications of existing ones have been suggested \cite{Newman2003, Boccaletti2006a}.

\subsection{Network characteristics}
\label{sec:network_characteristics}

In a binary network, the degree~$k_i$ of a node~$i$ is defined as the number of its neighbors ($k_i \defi \sum_{j=1}^n A_{ij}$). Its weighted counterpart is the strength $s_i \defi \sum_{j=1}^n W_{ij}$.
To simplify notation, we define $A_{ii} \defi 0 \,\forall\, i$ and $W_{ii} \defi 0 \,\forall\, i$ in the following.

The binary \textbf{clustering coefficient} $C_i$ of node~$i$ is defined as the rate, at which its neighbors are connected to each other:
\[ C_i \defi \frac{2}{ k_i \kl{k_i - 1} } \sum_{j=1}^n \sum_{l=1}^{j-1} A_{ij} A_{jl} A_{li} \]
Its mean over all nodes is the clustering coefficient~$C$ of the network.
Note that this is different from the rate at which nodes with a common neighbor are connected (transitivity), which is also used sometimes \cite{Stam2007b, vanDellen2009, Stam2010a}.

Several suggestions have been made on generalizing the clustering coefficient to weighted networks~\cite{Saramaki2007}.
Most of these may, however, be infeasible for application to complete weighted networks, since the clustering coefficient is always $1$ or irremovably discontinuous, if edge weights tend to~0~\cite{Saramaki2007}.

The clustering coefficient may be artificially increased for functional networks due to spatially close sensors picking up activities from common sources and due to the incapability of most estimators to distinguish between direct and indirect interactions \cite{Bialonski2010, Gerhard2011, Zalesky2012} (see also Section~\ref{sec:estimating_interactions}).

While the definition of the \textbf{shortest path}~$d_{ij}$ between two nodes in a binary network is straightforward, in a weighted network, the question arises how to define the ``length'' of a single edge.
Usually the inverse of the edge weight~$W_{ij}^{-1}$ is used.
Disconnected networks, i.e., networks with $d_{ij} = \infty$ for some $i$ and~$j$, pose a problem, since this would render the mean shortest path $L$ of the network to be~$\infty$.
This problem can be avoided by employing a network construction scheme that does not allow for unconnected networks (see also Section~\ref{sec:constructing_networks}).
Another way of dealing with unconnected networks is to regard the harmonic mean over the shortest paths~$d_{ij}$ instead of the arithmetic one, or more precisely its inverse, the \textit{efficiency.}
Shortest paths may be underestimated due to spurious shortcuts that may arise, e.g., from statistical fluctuations, common sources, and indirect interactions \cite{Ioannides2007, Bialonski2010}.

\textbf{Assortativity} quantifies whether nodes preferentially connect to nodes with a similar degree~\cite{Newman2002a}.
For binary networks, the assortativity $a$ is defined as the correlation coefficient over $\klg{\kl{k_i,k_j} \middle| A_{ij} = 1, 1\leq i,j \leq n}$:
\[\textstyle
a \defi
\kl{ 2 K_1 \sum_{i=1}^n \sum_{j=1}^{i-1} A_{ij} k_i k_j - K_2^2 }
/
\kl{K_1 K_3 - K_2^2} \]
with $K_m \defi \sum_{l=1}^n k_l^m$.
A generalization to weighted networks (using weighted statistics)~\cite{Leung2007} can be obtained by replacing $k$ with $s$ and $A$ with $W$.
Common sources (see Section \ref{sec:estimating_interactions}) can affect $a$, causing interaction networks to be classified as assortative ($a$ positive) even if the underlying interaction structure is dissortative ($a$ negative)~\cite{Bialonski2012}.

\textbf{Synchronizability}\label{synchronizability} describes the stability of the globally synchronized state of a network \cite{Barahona2002, Atay2006}.
With $\lambda_n$ denoting the largest eigenvalue of the Laplacian $\mathcal{L}$ of the network ($\mathcal{L}_{ij} \defi k_i \delta_{ij} - A_{ij}$, where $\delta$ is the Kronecker delta) and $\lambda_2$ denoting the second smallest (the smallest being~$0$), the \textit{eigenratio} of the network is defined as $S=\lambda_n / \lambda_2$.
Given some node dynamics, if $S$ exceeds a certain threshold, the synchronized state of the network is unstable.
Note that the terminology used in this context is highly inconsistent, e.g., ``synchronizability'' has been used as a name for $S$ \cite{Stam2012b} as well as for $S^{-1}$ \cite{vanWijk2010}.

\textbf{Centralities} \cite{Freeman1979, Bonacich1987, Koschutzki2005, Estrada2010} estimate the importance of a node in a network.
\textit{Degree centrality} $Z^\text{D}_i$ is defined as the degree or the strength of node $i$.
Nodes connected to many other nodes are called \textit{hubs} and are assumed to be more important than other nodes.

\textit{Closeness centrality} is defined as the inverse of the average of shortest paths between node $i$ and all other nodes in a network:
\[
	Z^\text{C}_i \defi \frac{n(n-1)}{\sum_{j=1}^{n} d_{ij}},
\]
with $Z^\text{C}_i \in [0,1]$.
For disconnected networks, $Z^\text{C}_i$ is zero for all nodes.
Alternatively, $Z^\text{C}_i$ can be estimated for the subnetwork a specific node is part of, by averaging over only those nodes that can be reached from that node.

\textit{Betweenness centrality} of node $i$ is defined as:
\[
	Z^\text{B}_i \defi \frac{2}{(n-1)(n-2)}
	\sum_{j=1, j \neq i}^{n}~
	\sum_{l=1, l \neq i}^{j-1}
	\frac{\eta_{jl}(i)}{H_{jl}},
\]
where $H_{jl}$ is the number of shortest paths between nodes $j$ and $l$, and $\eta_{jl}(i)$ is the number of these shortest paths that pass through node $i$.

\subsection{Normalizations and null models}\label{nullmodel}
When interpreting the aforementioned network characteristics, there are some factors whose influence one might consider spurious and want to eliminate.

For example, in most studies of binary functional networks, the total number of edges is considered spurious for two reasons:
first, it strongly depends on the threshold used for network construction (see Section \ref{sec:constructing_networks}).
Second, it reflects the average level of strength of interaction over all pairs of nodes, which may be a meaningful quantity, but can be assessed without the network approach.
The usual way to eliminate this influence is to take into account the expectation of the network characteristic under consideration for null-model networks which have the same number of edges as the original network, but are otherwise random.
For these Erd\H{o}s--R\'enyi random graphs~\cite{Erdos1959}, the expected value of some characteristics, e.g., the clustering coefficient, is known, while others have to be estimated analytically or via Monte Carlo-simulated instances of the null model \textit{(surrogates).}
For binary networks further methods are available to analytically or numerically obtain the expectation of characteristics for null models that also constrain the degrees \cite{Newman2001b, Maslov2004, ArtzyRandrup2005, Foster2007, DelGenio2010} or more complex properties~\cite{Annibale2009}.

Complete weighted networks are not influenced by a choice of threshold and the influence of the average level of strength of interaction can be eliminated during network construction (see Section \ref{sec:constructing_networks}).
However, there are other influencing factors, which one might consider spurious:
e.g., the weighted generalization of the clustering coefficient proposed in Ref.~\cite{Onnela2005} uses the maximum edge weight for normalization and may therefore be dominated by this quantity~\cite{Ansmann2011}.
But even without this normalization, this clustering coefficient and the mean shortest path length may be dominated by the standard deviation of the edge weights~\cite{Ansmann2011}.
As for binary networks these influences can be spotted and even eliminated by contrasting the characteristics for a network under investigation with those for respective null models.
For weighted networks, surrogates that preserve the weights~\cite{Barrat2004b} or the strengths \cite{Serrano2008, Ansmann2011} as well as analytical approaches~\cite{Garlaschelli2009} are available.

Recently, it has been proposed to also take into account the way networks are inferred and compare the results of the whole analysis procedure to those for null-model time series \cite{Bialonski2011b, Hlinka2012}.

\section{Characterizing evolving epileptic networks}
\label{sec:epinetworks}
Since the concept of an epileptic network comprises anatomically and functionally connected brain structures, in the following we report on and summarize findings obtained from studies that aim at assessing structural and functional network alterations associated with epilepsy.

\textbf{Structural alterations.} Structural imaging techniques such as magnetic resonance imaging (MRI) are widely used to identify cerebral lesions that might be associated with epilepsy~\cite{Duncan2010}.
MRI-based techniques such as voxel-based morphometry~\cite{Yasuda2010}, diffusion tensor imaging \cite{Gross2011, Bonilha2012, Vaessen2012}, or cortical thickness correlations~\cite{Bernhardt2011} have often revealed structural abnormalities that extend well beyond the possibly epileptogenic lesion, indicating widespread changes \cite{Guye2010, Richardson2010}.
These mostly subtle alterations in tissue integrity are thought to play a crucial role in epilepsy as they may significantly modify the global topology of structural and functional networks~\cite{Zhang2011b}.
These findings, however, have to be interpreted with care, given the many inconsistencies between studies \cite{Park2008, Richardson2010}.
Structural changes associated with epilepsy usually evolve on time scales much larger than the epileptic dynamics (epileptic discharges, seizures, etc.) and are thus assumed to be static phenomena.
It should be noted though that measurements repeated serially over time could in principle capture evolving structural changes related to, e.g., the natural history of the disease or treatment interventions.

\textbf{Functional alterations.} Several characteristics of evolving epileptic networks
have been assessed to improve understanding of seizure dynamics, which might help to answer the still unsolved issues of seizure initiation, spread, and termination in humans \cite{Mormann2007, Lado2008}.

Refs. \cite{Ponten2007, Schindler2008a, Ponten2009, Kramer2010, Kuhnert2010, Bialonski2011b, Gupta2011} report the respective null-model-normalized clustering coefficient ($\tilde C$) and the respective null-model-normalized mean shortest path length ($\tilde L$) to exhibit a concave-like temporal evolution during focal seizures (with or without secondary generalization), during absence seizures (primary generalized seizures), and during status epilepticus (which may be defined as non-terminating seizure activity characterized by epileptiform EEG patterns and concurrent behavioral changes~\cite{Chen2006}).
Interestingly, these consistent findings have been achieved despite the use of different estimators for the characterization of interactions.
When interpreted within the framework of the Watts--Strogatz model~\cite{Watts1998, Watts1999}, the evolution of network characteristics is indicative of a movement from a more random (before seizure; lower $\tilde C$ and $\tilde L$) toward a more regular (during seizure; higher $\tilde C$ and $\tilde L$) and then toward a more random (seizure ending and after seizure) functional topology.
For the 100 seizures from 60 patients investigated in Ref.~\cite{Schindler2008a}, the authors report the eigenratio (see Section \ref{synchronizability}) to significantly increase during the seizure but to decrease already prior to seizure end.
The decreased eigenratio may catalyze the emergence of a globally synchronized state of the epileptic brain, which could probably be regarded as a seizure-terminating mechanism \cite{Topolnik2003, Schiff2005, Schindler2007a, Schindler2007b, Kramer2010, Mueller2011}.
For the same data, Ref.~\cite{Bialonski2012} reports on a concave-like temporal evolution of assortativity, indicating that epileptic networks exhibit a more assortative topology during seizures than before or after seizures.
Positive assortativity values have recently been reported for other functional brain networks \cite{Jalili2011, Braun2012} and indicate that networks are likely to have a comparatively resilient core of mutually interconnected high-degree nodes~\cite{Newman2002a}.

There are by now only a few studies that investigated the temporal evolution of node-specific characteristics of epileptic networks during seizures \cite{Kramer2008, Wilke2011, Varotto2012}.
Findings range from an increased betweenness and degree centrality of few nodes at seizure onset to an evolution of the betweenness centrality of focal and non-focal nodes that starts more or less stable and decreases towards seizure end.
If highest values of centralities had been observed for focal nodes, these nodes have been interpreted as network hubs facilitating seizure activity.
Highest values of centralities were, however, not necessarily confined to the epileptic focus, and the results varied over the different centralities.
Similar findings have recently been obtained for the temporal evolution of degree, closeness, and betweenness centrality of nodes of functional brain networks from healthy subjects recorded during different physiological conditions~\cite{Kuhnert2012}.

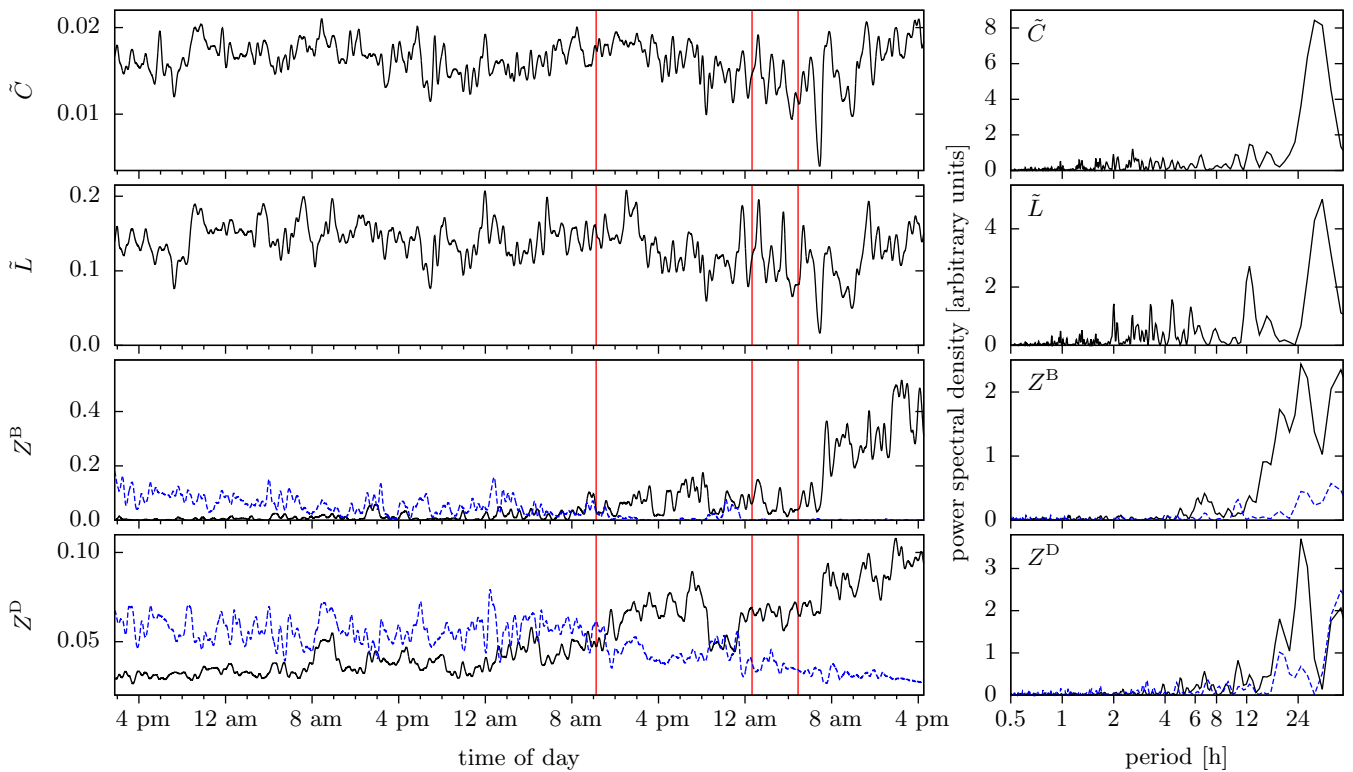
\begin{figure*}
	\input{Zeitreihen.tex}
	\caption{Exemplary temporal evolutions (left) and corresponding periodograms (right) of some characteristics of weighted networks with $n=72$ nodes constructed from iEEG data recorded over three days from an epilepsy patient (patient~9 in Ref.~\cite{Kuhnert2010}).
	The weight matrices were obtained via $W_{ij} = I_{ij}/\bar{I}$, where $I$ denotes the interaction matrix derived using the mean phase coherence (see Ref.~\cite{Kuhnert2010} for details).
	Red vertical lines mark electrical seizure onsets; seizures lasted for about half a minute.
	First row: temporal evolution of the surrogate-normalized clustering coefficient $\tilde{C} \defi \kl{C - C_0}/C_0$, where $C \defi \binom{n}{3}^{-1} \sum_{i=1}^n \sum_{j=1}^{i-1} \sum_{k=1}^{j-1} \sqrt[3]{W_{ij} W_{jk} W_{ki}}$ and $C_0$ is the mean over $C$ of 4096 weight-preserving surrogates~\cite{Barrat2004b}.
	Second row: the same for the normalized mean shortest path length $\tilde{L}$ (using $d_{ij}\defi W_{ij}^{-1}$).
	Third row: temporal evolution of the betweenness centrality $Z^\text{B}$ of a node associated with the epileptic focus (blue, dashed line) and of a node associated with a brain region distant from the epileptic focus (black, solid line).
	Fourth row: the same for the degree centrality $Z^\text{D}$.
	The variability of all characteristics is dominated by daily rhythms (around 24~hours; see the corresponding Lomb--Scargle periodograms on the right).
	Note the change of the order of node importance over time, assessed consistently by both centralities.
	All time series were smoothed using a Gaussian kernel with $\sigma = \unit[5]{min}$ for better legibility.
	}
	\label{fig:Hirnzeitreihen}
\end{figure*}

Findings that have been achieved so far for evolving epileptic networks during seizures appear quite intriguing, given the similarity of topological evolution across different types of epilepsies, seizures, medication, age, gender, and other clinical features.
This might point to comparable biophysical mechanisms underlying initiation, spread, and termination of focal and generalized seizures.
Nevertheless, improving our understanding of mechanisms underlying seizure generation also requires knowledge about characteristics of evolving epileptic networks during the seizure-free interval.
Several studies report on an altered functional brain topology in epilepsy patients even during the seizure-free interval when compared to healthy controls \cite{vanDellen2009, Chavez2010, Horstmann2010, Liao2010, Vlooswijk2011, Zhang2011b, Ansmann2012}.
However, findings have mostly been achieved from analyses of neural activities recorded over comparably short periods of time (a few tens of seconds up to a few minutes).
By now, the evolution of epileptic networks over longer periods (days to weeks) has been monitored in few studies only.
In Ref.~\cite{Kuhnert2010} evolving epileptic networks have been derived from long-term, multichannel iEEG data from 13~epilepsy patients.
The recording duration totaled more than \unit[2100]{h} (range: \unit[90--267]{h}) during which 75~seizures and one status epilepticus occurred.
For time-resolved estimates of mean shortest path lengths and clustering coefficients, large fluctuations over time could be observed, however, with some periodic temporal structure, which could be attributed --~to a large extent~-- to daily rhythms.
For some patients contributions on timescales longer than one day could be observed, which can possibly be attributed to changes of the anticonvulsant medication during presurgical evaluation.
Relevant aspects of the epileptic process acting on timescales from seconds up to a few hours (such as clinical and subclinical seizures, possible seizure precursors, epileptiform activities during the seizure-free interval, as well as their interferences with physiological processes) contributed to a much lesser extent to the temporal variability of the network characteristics (cf.~Fig.~\ref{fig:Hirnzeitreihen}).
Interestingly, despite being less pronounced, alterations of the network characteristics during the pre-seizure period pointed to a loss of functional long-range connections prior to seizures.
It remains to be shown if these alterations allow for unequivocally judging whether a more random network configuration promotes seizure generation, as proposed earlier in Refs.~\cite{Ponten2007, Ponten2009}.

Large fluctuations of network characteristics have also been observed in another study~\cite{Kramer2011} that investigated evolving epileptic networks derived from iEEG data recorded from six patients over periods of approximately \unit[24]{h}.
Time-resolved analyses of several network characteristics revealed sparse, fractured, and modular network topologies with large temporal variability.
Within this variability the authors observed the emergence of persistent structures (probability of edge appearance in time) on short timescales (approx.~\unit[100]{s}) with high consistency across cognitive states and days of recording.
Similar findings have recently been reported for evolving brain networks of healthy subjects~\cite{Chu2012}.

Our preliminary findings (Fig.~\ref{fig:Hirnzeitreihen}) indicate that node-specific characteristics (e.g., centralities) also exhibit large long-term fluctuations, which might reflect daily rhythms.
These findings also indicate that the epileptic focus is not consistently the most important node, but importance may drastically vary over time.
This is in contrast to previous studies \cite{Wilke2011, Varotto2012} that reported on highest centrality values for the epileptic focus, which might be due to considering short-lasting (some tens of seconds to a few minutes) iEEG recordings only.

\section{Conclusion and Outlook}
\label{sec:conclusion}

We summarized recent conceptual and methodological developments aiming at an improved inference and characterization of human epileptic brain networks and reported on findings obtained so far for evolving epileptic networks on timescales ranging from a few seconds to days and weeks.
In a work of this scope it is inevitable that some contributions may be over- or underemphasized, depending upon the points to be made in the text.

There is converging evidence for properties of evolving brain networks during epileptic seizures to indicate --~when interpreted within the framework of the Watts--Strogatz model~\cite{Watts1998, Watts1999}~-- a movement from a more random toward a more regular and back toward a more random functional topology, with highest resilience and least stability of the synchronous state during the more regular topology.
These findings may provide clues as to how seizures self-terminate and as to how to control epileptic networks, which may fertilize research into alternative therapeutic options.
During the seizure-free interval, large fluctuations of network- and node-specific properties of epileptic brain networks have been observed recently that, to a large extent, can be attributed to daily rhythms.
Since relevant aspects of the epileptic process appear to contribute only marginally to these fluctuations, the entanglement of physiological and pathophysiological dynamics may affect properties of evolving brain networks and thus be a confounding variable that hinders progress in improving our understanding of seizure generation in epileptic brain networks.
Clearly, more work is necessary to better understand the dynamics of the seizure-free interval.

Additional insights might be achieved from computational models of epilepsy \cite{Lytton2008, Soltesz2008, Wendling2008}.
Results from a number of simulation studies \cite{Buzsaki2004b, Netoff2004, Roxin2004, Percha2005, Feldt2007, Riecke2007, Morgan2008, Rothkegel2009, Raiesdana2011, Rothkegel2011, Anderson2012, Benjamin2012, Rothkegel2012}  already emphasize the crucial role of complex network topologies for the initiation, spread, and termination of seizure-like activity and point to intricate interactions between network structure and intrinsic properties of neurons \cite{Dyhrfjeld-Johnsen2007, Bogaard2009}, which, however, remain poorly understood.
Progress along this line can be expected from recently developed recording techniques that allow sampling of neural activities in humans with high temporal resolution on sub-millimeter spatial scales \cite{Petridou2006, Zotev2008, Stead2010, Pan2011, Truccolo2011, Viventi2011, Bower2012}.

Inferring evolving brain networks from multichannel recordings of neural activities is notoriously difficult, given the challenges associated with defining nodes in spatially extended dynamics systems \cite{Butts2009, Bialonski2010} together with the challenges associated with defining edges from interaction estimates (such as indirect interactions, common sources, multiple timescales, as well as statistical and robustness issues).
Despite the availability of a number of improved or newly developed bivariate and multivariate time series analysis techniques, it remains to be investigated whether these techniques or other methods to map between time series and complex networks \cite{Zhang2006a, Lacasa2008, Gao2009, Donner2010, Campanharo2011, Emmert-Streib2011, Iwayama2012, Nakamura2012} allow for an improved network inference.
Last but not least, there is a strong need for improved network characteristics and null models (and corresponding surrogates), allowing for a more robust classification of weighted and/or directed evolving brain networks, and for statistical tests allowing for a robust comparison of networks across studies together with reliable estimates for the significance of findings.

We are confident that further developments will allow for an improved inference and characterization of evolving epileptic brain networks, which will advance the understanding of the dynamical disease epilepsy and may guide new developments for diagnosis, treatment, and control.

\section*{Acknowledgements}
This work was supported by the Deutsche Forschungsgemeinschaft (Grants Nos.~LE660/4\nobreakdash-2 and LE660/5\nobreakdash-2).

\input{Manuscript.bbl}

\end{document}

%% file: Zeitreihen.tex
\begingroup

  \makeatletter
  \providecommand\color[2][]{%
    \GenericError{(gnuplot) \space\space\space\@spaces}{%
      Package color not loaded in conjunction with
      terminal option `colourtext'%
    }{See the gnuplot documentation for explanation.%
    }{Either use 'blacktext' in gnuplot or load the package
      color.sty in LaTeX.}%
    \renewcommand\color[2][]{}%
  }%
  \providecommand\includegraphics[2][]{%
    \GenericError{(gnuplot) \space\space\space\@spaces}{%
      Package graphicx or graphics not loaded%
    }{See the gnuplot documentation for explanation.%
    }{The gnuplot epslatex terminal needs graphicx.sty or graphics.sty.}%
    \renewcommand\includegraphics[2][]{}%
  }%
  \providecommand\rotatebox[2]{#2}%
  \@ifundefined{ifGPcolor}{%
    \newif\ifGPcolor
    \GPcolortrue
  }{}%
  \@ifundefined{ifGPblacktext}{%
    \newif\ifGPblacktext
    \GPblacktexttrue
  }{}%
  \let\gplgaddtomacro\g@addto@macro
  \gdef\gplbacktext{}%
  \gdef\gplfronttext{}%
  \makeatother
  \ifGPblacktext
    \def\colorrgb#1{}%
    \def\colorgray#1{}%
  \else
    \ifGPcolor
      \def\colorrgb#1{\color[rgb]{#1}}%
      \def\colorgray#1{\color[gray]{#1}}%
      \expandafter\def\csname LTw\endcsname{\color{white}}%
      \expandafter\def\csname LTb\endcsname{\color{black}}%
      \expandafter\def\csname LTa\endcsname{\color{black}}%
      \expandafter\def\csname LT0\endcsname{\color[rgb]{1,0,0}}%
      \expandafter\def\csname LT1\endcsname{\color[rgb]{0,1,0}}%
      \expandafter\def\csname LT2\endcsname{\color[rgb]{0,0,1}}%
      \expandafter\def\csname LT3\endcsname{\color[rgb]{1,0,1}}%
      \expandafter\def\csname LT4\endcsname{\color[rgb]{0,1,1}}%
      \expandafter\def\csname LT5\endcsname{\color[rgb]{1,1,0}}%
      \expandafter\def\csname LT6\endcsname{\color[rgb]{0,0,0}}%
      \expandafter\def\csname LT7\endcsname{\color[rgb]{1,0.3,0}}%
      \expandafter\def\csname LT8\endcsname{\color[rgb]{0.5,0.5,0.5}}%
    \else
      \def\colorrgb#1{\color{black}}%
      \def\colorgray#1{\color[gray]{#1}}%
      \expandafter\def\csname LTw\endcsname{\color{white}}%
      \expandafter\def\csname LTb\endcsname{\color{black}}%
      \expandafter\def\csname LTa\endcsname{\color{black}}%
      \expandafter\def\csname LT0\endcsname{\color{black}}%
      \expandafter\def\csname LT1\endcsname{\color{black}}%
      \expandafter\def\csname LT2\endcsname{\color{black}}%
      \expandafter\def\csname LT3\endcsname{\color{black}}%
      \expandafter\def\csname LT4\endcsname{\color{black}}%
      \expandafter\def\csname LT5\endcsname{\color{black}}%
      \expandafter\def\csname LT6\endcsname{\color{black}}%
      \expandafter\def\csname LT7\endcsname{\color{black}}%
      \expandafter\def\csname LT8\endcsname{\color{black}}%
    \fi
  \fi
  \setlength{\unitlength}{0.0500bp}%
  \begin{picture}(10080.00,5760.00)%
    \gplgaddtomacro\gplbacktext{%
      \csname LTb\endcsname%
      \put(700,996){\makebox(0,0)[r]{\strut{}0.05}}%
      \put(700,1669){\makebox(0,0)[r]{\strut{}0.10}}%
      \put(989,398){\makebox(0,0){\strut{} 4 pm}}%
      \put(1642,398){\makebox(0,0){\strut{}12 am}}%
      \put(2294,398){\makebox(0,0){\strut{} 8 am}}%
      \put(2947,398){\makebox(0,0){\strut{} 4 pm}}%
      \put(3600,398){\makebox(0,0){\strut{}12 am}}%
      \put(4252,398){\makebox(0,0){\strut{} 8 am}}%
      \put(4905,398){\makebox(0,0){\strut{} 4 pm}}%
      \put(5557,398){\makebox(0,0){\strut{}12 am}}%
      \put(6210,398){\makebox(0,0){\strut{} 8 am}}%
      \put(6863,398){\makebox(0,0){\strut{} 4 pm}}%
      \put(128,1198){\rotatebox{-270}{\makebox(0,0){\strut{}$Z^\text{D}$}}}%
      \put(3855,112){\makebox(0,0){\strut{}time of day}}%
    }%
    \gplgaddtomacro\gplfronttext{%
    }%
    \gplgaddtomacro\gplbacktext{%
      \csname LTb\endcsname%
      \put(700,1912){\makebox(0,0)[r]{\strut{}0.0}}%
      \put(700,2322){\makebox(0,0)[r]{\strut{}0.2}}%
      \put(700,2732){\makebox(0,0)[r]{\strut{}0.4}}%
      \put(989,1717){\makebox(0,0){\strut{}}}%
      \put(1642,1717){\makebox(0,0){\strut{}}}%
      \put(2294,1717){\makebox(0,0){\strut{}}}%
      \put(2947,1717){\makebox(0,0){\strut{}}}%
      \put(3600,1717){\makebox(0,0){\strut{}}}%
      \put(4252,1717){\makebox(0,0){\strut{}}}%
      \put(4905,1717){\makebox(0,0){\strut{}}}%
      \put(5557,1717){\makebox(0,0){\strut{}}}%
      \put(6210,1717){\makebox(0,0){\strut{}}}%
      \put(6863,1717){\makebox(0,0){\strut{}}}%
      \put(129,2516){\rotatebox{-270}{\makebox(0,0){\strut{}$Z^\text{B}$}}}%
    }%
    \gplgaddtomacro\gplfronttext{%
    }%
    \gplgaddtomacro\gplbacktext{%
      \csname LTb\endcsname%
      \put(700,3231){\makebox(0,0)[r]{\strut{}0.0}}%
      \put(700,3793){\makebox(0,0)[r]{\strut{}0.1}}%
      \put(700,4356){\makebox(0,0)[r]{\strut{}0.2}}%
      \put(989,3036){\makebox(0,0){\strut{}}}%
      \put(1642,3036){\makebox(0,0){\strut{}}}%
      \put(2294,3036){\makebox(0,0){\strut{}}}%
      \put(2947,3036){\makebox(0,0){\strut{}}}%
      \put(3600,3036){\makebox(0,0){\strut{}}}%
      \put(4252,3036){\makebox(0,0){\strut{}}}%
      \put(4905,3036){\makebox(0,0){\strut{}}}%
      \put(5557,3036){\makebox(0,0){\strut{}}}%
      \put(6210,3036){\makebox(0,0){\strut{}}}%
      \put(6863,3036){\makebox(0,0){\strut{}}}%
      \put(129,3835){\rotatebox{-270}{\makebox(0,0){\strut{}$\tilde{L}$}}}%
    }%
    \gplgaddtomacro\gplfronttext{%
    }%
    \gplgaddtomacro\gplbacktext{%
      \csname LTb\endcsname%
      \put(700,4975){\makebox(0,0)[r]{\strut{}0.01}}%
      \put(700,5628){\makebox(0,0)[r]{\strut{}0.02}}%
      \put(989,4355){\makebox(0,0){\strut{}}}%
      \put(1642,4355){\makebox(0,0){\strut{}}}%
      \put(2294,4355){\makebox(0,0){\strut{}}}%
      \put(2947,4355){\makebox(0,0){\strut{}}}%
      \put(3600,4355){\makebox(0,0){\strut{}}}%
      \put(4252,4355){\makebox(0,0){\strut{}}}%
      \put(4905,4355){\makebox(0,0){\strut{}}}%
      \put(5557,4355){\makebox(0,0){\strut{}}}%
      \put(6210,4355){\makebox(0,0){\strut{}}}%
      \put(6863,4355){\makebox(0,0){\strut{}}}%
      \put(128,5154){\rotatebox{-270}{\makebox(0,0){\strut{}$\tilde{C}$}}}%
    }%
    \gplgaddtomacro\gplfronttext{%
    }%
    \gplgaddtomacro\gplbacktext{%
      \csname LTb\endcsname%
      \put(7454,593){\makebox(0,0)[r]{\strut{}0}}%
      \put(7454,911){\makebox(0,0)[r]{\strut{}1}}%
      \put(7454,1230){\makebox(0,0)[r]{\strut{}2}}%
      \put(7454,1548){\makebox(0,0)[r]{\strut{}3}}%
      \put(7560,398){\makebox(0,0){\strut{}0.5}}%
      \put(7948,398){\makebox(0,0){\strut{}1}}%
      \put(8336,398){\makebox(0,0){\strut{}2}}%
      \put(8723,398){\makebox(0,0){\strut{}4}}%
      \put(8950,398){\makebox(0,0){\strut{}6}}%
      \put(9111,398){\makebox(0,0){\strut{}8}}%
      \put(9338,398){\makebox(0,0){\strut{}12}}%
      \put(9726,398){\makebox(0,0){\strut{}24}}%
      \put(7498,1198){\rotatebox{-270}{\makebox(0,0){\strut{}}}}%
      \put(8812,112){\makebox(0,0){\strut{}period [h]}}%
      \put(7166,3176){\rotatebox{-270}{\makebox(0,0){\strut{}\small power spectral density [arbitrary units]}}}%
      \put(7685,1622){\makebox(0,0)[l]{\strut{}$Z^\text{D}$}}%
    }%
    \gplgaddtomacro\gplfronttext{%
    }%
    \gplgaddtomacro\gplbacktext{%
      \csname LTb\endcsname%
      \put(7454,1912){\makebox(0,0)[r]{\strut{}0}}%
      \put(7454,2396){\makebox(0,0)[r]{\strut{}1}}%
      \put(7454,2879){\makebox(0,0)[r]{\strut{}2}}%
      \put(7560,1717){\makebox(0,0){\strut{}}}%
      \put(7948,1717){\makebox(0,0){\strut{}}}%
      \put(8336,1717){\makebox(0,0){\strut{}}}%
      \put(8723,1717){\makebox(0,0){\strut{}}}%
      \put(8950,1717){\makebox(0,0){\strut{}}}%
      \put(9111,1717){\makebox(0,0){\strut{}}}%
      \put(9338,1717){\makebox(0,0){\strut{}}}%
      \put(9726,1717){\makebox(0,0){\strut{}}}%
      \put(7498,2516){\rotatebox{-270}{\makebox(0,0){\strut{}}}}%
      \put(7685,2940){\makebox(0,0)[l]{\strut{}$Z^\text{B}$}}%
    }%
    \gplgaddtomacro\gplfronttext{%
    }%
    \gplgaddtomacro\gplbacktext{%
      \csname LTb\endcsname%
      \put(7454,3231){\makebox(0,0)[r]{\strut{}0}}%
      \put(7454,3671){\makebox(0,0)[r]{\strut{}2}}%
      \put(7454,4110){\makebox(0,0)[r]{\strut{}4}}%
      \put(7560,3036){\makebox(0,0){\strut{}}}%
      \put(7948,3036){\makebox(0,0){\strut{}}}%
      \put(8336,3036){\makebox(0,0){\strut{}}}%
      \put(8723,3036){\makebox(0,0){\strut{}}}%
      \put(8950,3036){\makebox(0,0){\strut{}}}%
      \put(9111,3036){\makebox(0,0){\strut{}}}%
      \put(9338,3036){\makebox(0,0){\strut{}}}%
      \put(9726,3036){\makebox(0,0){\strut{}}}%
      \put(7498,3835){\rotatebox{-270}{\makebox(0,0){\strut{}}}}%
      \put(7685,4259){\makebox(0,0)[l]{\strut{}$\tilde{L}$}}%
    }%
    \gplgaddtomacro\gplfronttext{%
    }%
    \gplgaddtomacro\gplbacktext{%
      \csname LTb\endcsname%
      \put(7454,4550){\makebox(0,0)[r]{\strut{}0}}%
      \put(7454,4819){\makebox(0,0)[r]{\strut{}2}}%
      \put(7454,5087){\makebox(0,0)[r]{\strut{}4}}%
      \put(7454,5356){\makebox(0,0)[r]{\strut{}6}}%
      \put(7454,5625){\makebox(0,0)[r]{\strut{}8}}%
      \put(7560,4355){\makebox(0,0){\strut{}}}%
      \put(7948,4355){\makebox(0,0){\strut{}}}%
      \put(8336,4355){\makebox(0,0){\strut{}}}%
      \put(8723,4355){\makebox(0,0){\strut{}}}%
      \put(8950,4355){\makebox(0,0){\strut{}}}%
      \put(9111,4355){\makebox(0,0){\strut{}}}%
      \put(9338,4355){\makebox(0,0){\strut{}}}%
      \put(9726,4355){\makebox(0,0){\strut{}}}%
      \put(7498,5154){\rotatebox{-270}{\makebox(0,0){\strut{}}}}%
      \put(7685,5578){\makebox(0,0)[l]{\strut{}$\tilde{C}$}}%
    }%
    \gplgaddtomacro\gplfronttext{%
    }%
    \gplbacktext
    \put(0,0){\includegraphics{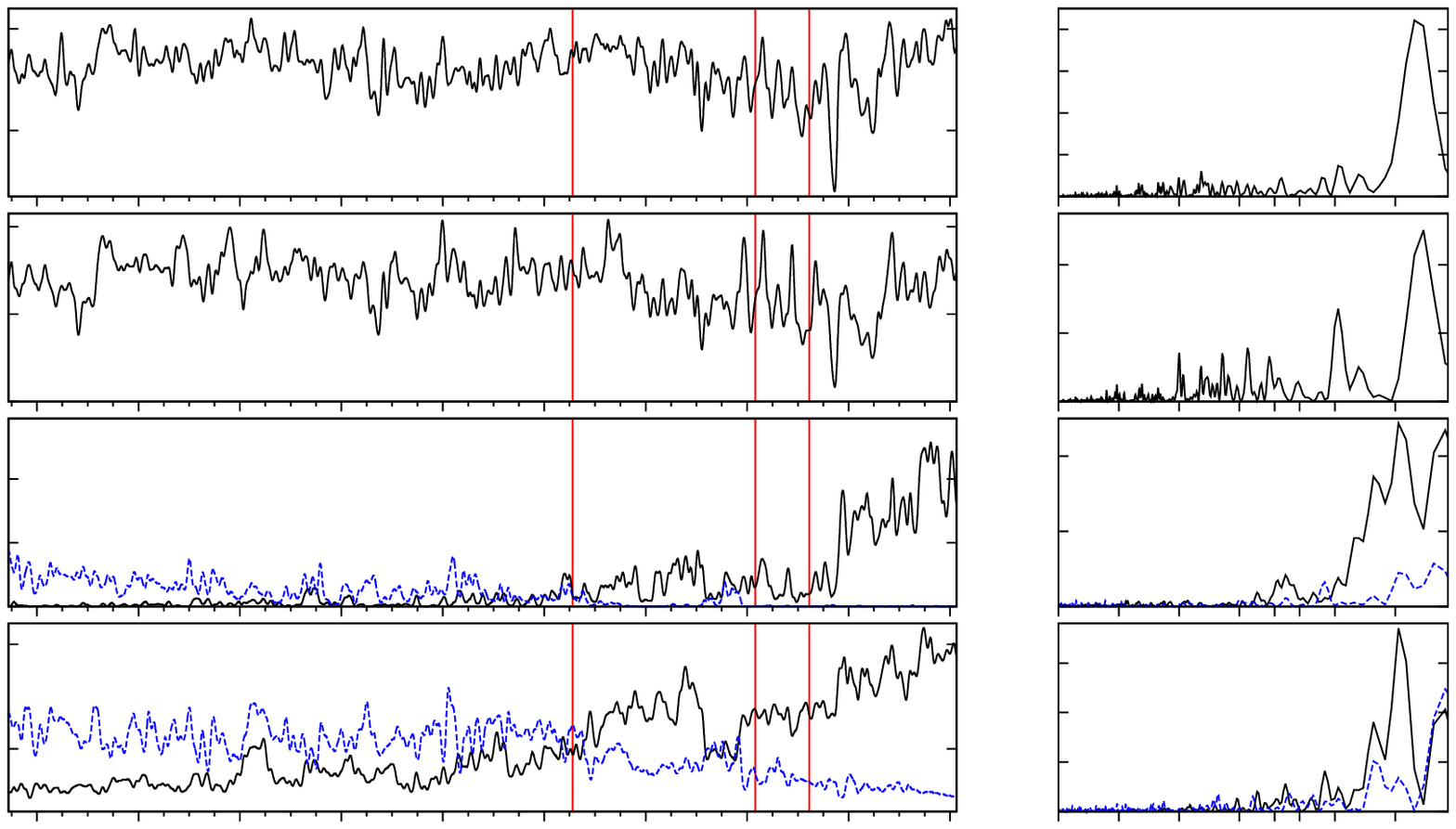}}%
    \gplfronttext
  \end{picture}%
\endgroup

%% file: Manuscript.bbl
%